\ifdefined\pdfximage\PassOptionsToPackage{pdftex}{graphicx}\fi
\documentclass[shortnote]{jpsj3}
\usepackage{txfonts}
\usepackage{bm}
\DeclareGraphicsExtensions{.eps,.pdf,.png,.jpg}

\title{Monte Carlo Study of the Phase Transition of the
$XY$ Model on a Diamond Lattice}

\author{Sena Watanabe$^1$, Yukitoshi Motome$^1$, and Haruki Watanabe$^{2,3,4,1}$\thanks{hwatanabe@ust.hk}}
\inst{$^1$Department of Applied Physics, The University of Tokyo, Tokyo 113-8656, Japan \\
$^2$Department of Physics, Hong Kong University of Science and Technology, Clear Water Bay, Hong Kong, China \\
$^3$Institute for Advanced Study, Hong Kong University of Science and Technology, Clear Water Bay, Hong Kong, China \\
$^4$Center for Theoretical Condensed Matter Physics, Hong Kong University of Science and Technology, Clear Water Bay, Hong Kong, China}

\abst{We study the phase transition of the classical $XY$ model on a
diamond lattice by Monte Carlo simulations using the Wolff cluster
algorithm.
Finite-size scaling (FSS) analysis of the Binder cumulant and
the second-moment correlation length ratio $\xi_{2\rm nd}/L$ yields
$T_c = 1.30036(1)$ and $\nu = 0.671(6)$.
Data collapse of both quantities confirms the
three-dimensional $XY$ universality class.}

\begin{document}
\maketitle

The classical $XY$ model---a system of two-component unit vectors with
nearest-neighbor interaction---is one of the most fundamental models in
statistical mechanics, exhibiting the
Berezinskii--Kosterlitz--Thouless transition in two dimensions and a
continuous transition to long-range order in three dimensions (3D).
While the critical behavior of the 3D transition has been extensively
studied on cubic lattices, investigations on other lattice geometries
remain scarce.
The $XY$ model on a diamond lattice has recently attracted interest
in the context of multipolar order in Pr-based 1-2-20
compounds\cite{Hattori2016} and as the dual representation of the
$S=1$ pyrochlore spin ice in the monopole-free
limit.\cite{Watanabe2026}
Hattori and Tsunetsugu studied this model with $Z_3$ single-ion
anisotropy, finding a continuous 3D $XY$
transition,\cite{Hattori2016} but a precise value of the
critical temperature $T_c$ of the isotropic $XY$ model
is still lacking, apart from an early high-temperature
series expansion study.\cite{Lambeth1974}
This Note provides a precise determination of $T_c$ for the isotropic
case and confirms the 3D $XY$ universality class.

\textit{Model and method.}---
We consider the classical ferromagnetic $XY$ model on the diamond lattice,
\begin{equation}
  H = -J \sum_{\langle i,j \rangle} \cos(\theta_i - \theta_j),
  \label{eq:ham}
\end{equation}
with $J=1$, where $\theta_i \in [0,2\pi)$ and the sum is over nearest-neighbor
pairs.
Since the diamond lattice is bipartite, a spin rotation by $\pi$ on one
sublattice maps this model exactly onto the antiferromagnetic one, so the
two cases share the same critical behavior.

The diamond lattice is a face-centered-cubic Bravais lattice with a
two-atom basis.
We place it in an $L^3$ simulation box using the standard
8-site conventional cubic cell, which contains four A~sites at the
face-centered-cubic positions and four B~sites displaced by
$\frac{1}{4}(1,1,1)$, giving $N = 8L^3$ sites with periodic boundary
conditions.
Each A~site has $z=4$ nearest neighbors on the B~sublattice with bond
vectors
$\bm{\delta}_k \in
\tfrac{1}{4}\bigl\{(1,1,1),\,(1,\bar{1},\bar{1}),\,
(\bar{1},1,\bar{1}),\,(\bar{1},\bar{1},1)\bigr\}$
($\bar{1} \equiv -1$).
The tetrahedral symmetry yields
$\sum_k \bm{\delta}_k = \bm{0}$ and
$\sum_k \delta_{k,\alpha}\delta_{k,\beta}\propto\delta_{\alpha\beta}$,
ensuring that the acoustic-mode dispersion is isotropic and the effective
continuum geometry of the simulation box is cubic---a prerequisite for
the dimensionless finite-size scaling (FSS) ratios to converge to the
known cubic universal values.\cite{Hasenbusch2019}

We simulate the model using the Wolff single-cluster
algorithm.\cite{Wolff1989}
A random reflection axis $\alpha\in[0,\pi)$ is drawn, and neighboring spins
are added to the cluster with probability
\begin{equation}
  p_{ij} = \max\bigl(0,\,
  1-e^{-2\beta J\sin(\theta_i-\alpha)\sin(\theta_j-\alpha)}\bigr),
  \label{eq:wolff}
\end{equation}
where $\beta = 1/T$; all cluster spins are then reflected across $\alpha$.
This algorithm practically eliminates critical slowing down,
keeping the integrated autocorrelation time
$\tau_{\rm int}\approx 1$ Wolff sweep for all system sizes.

Simulations were carried out for fourteen system sizes from
$L = 4$ to $56$ ($N$ from $512$ to $1{,}404{,}928$) on a 21-point temperature
grid $T \in [1.290, 1.310]$ with step $\Delta T = 0.001$.
For the FSS collapse analysis, additional ``wing'' temperatures extending
to $|T - T_c| \lesssim 0.1$ were computed for $L = 10$--$16$.
The number of measurements per temperature point was graded from
$n_{\rm meas} = 25{,}000$ at $L = 4$ to approximately $925{,}000$ at
$L = 56$.
All configurations were initialized via sequential cooling from $T = 1.45$
with $300$ Wolff sweeps per temperature, followed by an additional
thermalization of $\max(2000, 500L)$ Wolff sweeps before
measurement.\cite{notePhase}
Statistical errors were estimated by the jackknife method with 50 bins.

The primary observables are the Binder cumulant\cite{Binder1981}
\begin{equation}
  B = 1 - \frac{\langle Q^4\rangle}{2\langle Q^2\rangle^2},
  \label{eq:binder}
\end{equation}
where $Q = |\bm{Q}|$ with $\bm{Q} = N^{-1}\sum_i \bm{S}_i$
the magnetization and $\bm{S}_i = (\cos\theta_i, \sin\theta_i)$,
and the second-moment correlation length ratio $\xi_{2\rm nd}/L$, where
\begin{equation}
  \xi_{2\rm nd}
  = \frac{1}{2\sin(\pi/L)}
    \sqrt{\frac{S(\bm{0})}{S(\bm{q}_{\rm min})} - 1\,}
  \label{eq:xi2nd}
\end{equation}
with $S(\bm{q})
= N^{-1}\bigl\langle\bigl|\sum_i \bm{S}_i\,
e^{i\bm{q}\cdot\bm{r}_i}\bigr|^2\bigr\rangle$ the structure factor
and $\bm{q}_{\rm min} = (2\pi/L,0,0)$; the ratio $\xi_{2\rm nd}/L$
is averaged over the three Cartesian directions.
Both $B$ and $\xi_{2\rm nd}/L$ are dimensionless and approach universal
$L$-independent values at $T_c$ in the thermodynamic limit.

\begin{figure}[t]
\includegraphics[width=8.6cm]{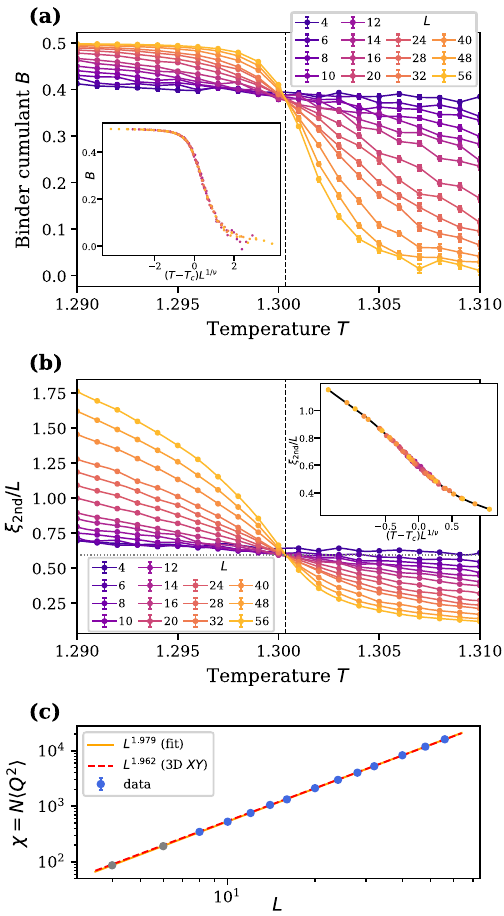}
\caption{(Color online)
(a) Binder cumulant $B$ vs $T$ for $L = 4$--$56$.
Inset: FSS collapse of $B$ for $L = 14$--$56$.
(b) $\xi_{2\rm nd}/L$ vs $T$ for $L = 4$--$56$.
Inset: FSS data collapse of $\xi_{2\rm nd}/L$ for $L = 12$--$56$,
$T \in [1.297, 1.303]$, with the polynomial fit (solid curve).
Vertical dashed lines in (a), (b): $T_c = 1.30036$;
horizontal dotted line in (b): 3D $XY$ universal value
$(\xi_{2\rm nd}/L)^* = 0.5927$.
(c) Susceptibility $\chi = N\langle Q^2\rangle$ vs $L$ at
$T_c = 1.30036$ (cubic-spline interpolation).
Solid line: power-law fit ($\gamma/\nu = 1.98$);
dashed line: accepted 3D $XY$ slope ($\gamma/\nu = 1.962$).
Gray points ($L < 8$) are excluded from the fit.}
\label{fig:main}
\end{figure}

\textit{Numerical results.}---
Figure~\ref{fig:main}(a) shows the Binder cumulant $B$ versus $T$ for
all fourteen system sizes.
The curves from different system sizes cross near $T \approx 1.3$,
signaling a continuous phase transition.
A polynomial FSS fit of $B$ against the scaling variable
$x = (T - T_c)\,L^{1/\nu}$ for $L \ge 12$ in the range
$T \in [1.297, 1.303]$ yields $T_c \approx 1.3004$ and
$\nu \approx 0.69$, with the reduced chi-squared
$\chi^2_{\rm red} = 2.0$,
consistent with the 3D $XY$ value
$\nu = 0.6717(1)$.\cite{Hasenbusch2019}
However, the ratio $\langle Q^4\rangle/\langle Q^2\rangle^2$ in
$B$ amplifies statistical noise in the disordered phase where
$\langle Q^2\rangle$ becomes small, limiting the achievable precision.

To refine the critical temperature, we analyze $\xi_{2\rm nd}/L$
[Fig.~\ref{fig:main}(b)], which does not suffer from this noise
amplification.
The size-doubling ($b = 2$) crossing temperatures for
the seven largest pairs ($L_1 = 10$--$28$) cluster tightly within
$1.3001$--$1.3005$, indicating rapid convergence.
A polynomial FSS fit $\xi_{2\rm nd}/L = \sum_{n=0}^{5} a_n x^n$
with $x = (T - T_c)\,L^{1/\nu}$
to the data for $L \ge 12$ in the range $T \in [1.297, 1.303]$
[inset of Fig.~\ref{fig:main}(b)] yields
\begin{equation}
  T_c = 1.30036 \pm 0.00001, \quad
  \nu = 0.671 \pm 0.006,
  \label{eq:Tc}
\end{equation}
with $\chi^2_{\rm red} = 1.32$ ($70$ data points, $8$ parameters).
The improvement over the Binder analysis is reflected in
both the smaller uncertainties and the lower $\chi^2_{\rm red}$.
Fixing $\nu = 0.6717$ gives $T_c = 1.30036(1)$ with
$\chi^2_{\rm red} = 1.30$; the 3D $XY$ value lies within the $1\sigma$
region of the joint $(T_c, \nu)$ confidence contour.

Corrections to scaling affect $\nu$ but not $T_c$.
A pure-scaling collapse gives an effective $\nu$ drifting from $\approx 0.66$
to $\approx 0.68$ as the smallest fitted size is raised; including a leading
correction $(1 + c\,L^{-\omega})$ with $\omega = 0.789$\cite{Hasenbusch2019}
stabilizes $\nu = 0.670(6)$, while fixing $(\xi_{2\rm nd}/L)^* = 0.5927$
returns $T_c = 1.30036(1)$ independently of the fit range.

The crossing values $(\xi_{2\rm nd}/L)^*$ from the seven largest
pairs lie in the range $0.577$--$0.601$,
within $1$--$3\%$ of the 3D $XY$ cubic universal value
$(\xi_{2\rm nd}/L)^* = 0.5927$,\cite{Hasenbusch2019} with the
residual deviation attributable to corrections to scaling
($\omega = 0.789$).\cite{Hasenbusch2019}
The clean data collapse of both $B$ and $\xi_{2\rm nd}/L$
[insets of Fig.~\ref{fig:main}(a) and (b)] confirms the
3D $XY$ universality class.

To extract other critical exponents, we analyze finite-size power-law behavior at $T_c$. 
Physical quantities at $T_c$ are obtained by cubic-spline interpolation of
the temperature-grid data to $T_c = 1.30036$; this eliminates the systematic
bias that would arise from evaluating observables at the nearest grid point
$T = 1.300$, where the finite scaling variable
$x = (T - T_c) L^{1/\nu} \approx -0.06$ for $L = 32$
shifts the effective exponents appreciably.
We fit the susceptibility
\begin{equation}
  \chi = N\langle Q^2\rangle \sim L^{\gamma/\nu}
  \label{eq:chi_scal}
\end{equation}
for $L = 8$--$56$ [Fig.~\ref{fig:main}(c)].
Since $\chi = 8L^3 \langle Q^2\rangle$, the order-parameter
exponent ratio $2\beta/\nu = d - \gamma/\nu$ follows identically from the
same data and is not an independent fit.
Na\"{\i}ve log-log fits give effective exponents
$\gamma/\nu \approx 1.98$ and $2\beta/\nu \approx 1.02$,
in good agreement with the 3D $XY$ values
$\gamma/\nu = 1.962$ and
$2\beta/\nu = 1.038$.\cite{Campostrini2001,Hasenbusch2019}
The residual ${\sim}\,1\%$ deviation is attributable to corrections
to scaling of order $L^{-\omega}$ with
$\omega = 0.789$;\cite{Hasenbusch2019}
the dashed line in Fig.~\ref{fig:main}(c) shows the accepted
3D $XY$ slope $\gamma/\nu = 1.962$ for comparison.

\textit{Conclusion.}---
Using the Wolff cluster algorithm on system sizes up to $L=56$
($N = 1{,}404{,}928$), we have precisely determined the critical temperature
of the isotropic $XY$ model on the diamond lattice:
$T_c = 1.30036(1)$, approximately $2.4\%$ above the anisotropic
value $T_c\approx 1.2695(3)$ obtained with $Z_3$ single-ion
anisotropy.\cite{Hattori2016}
The reduction of $T_c$ by the $Z_3$ anisotropy can be understood by
noting that, under the sublattice rotation that maps the
antiferromagnet onto a ferromagnet, a uniform single-ion
anisotropy becomes sublattice-dependent and frustrates uniform ordering,
thereby lowering $T_c$.
FSS analysis of both the Binder cumulant and
the correlation length ratio $\xi_{2\rm nd}/L$ yields the critical
exponent $\nu = 0.671(6)$, consistent with the 3D $XY$ value
$\nu = 0.6717(1)$,\cite{Hasenbusch2019} and the data collapse of
both quantities confirms the 3D $XY$ universality class.
Power-law fits of the susceptibility at $T_c$ yield
$\gamma/\nu = 1.98$ and $2\beta/\nu = 1.02$, consistent with the
3D $XY$ values to within $\sim 1\%$.
These results agree with, and substantially sharpen, the early
tenth-order high-temperature series estimates of Lambeth
\textit{et al.}\cite{Lambeth1974} for the planar model on the diamond lattice:
$\gamma = 1.34(2)$ and $2\nu+\gamma = 2.70(5)$ give $\gamma/\nu \approx 1.97$
and $\nu \approx 0.68$, and their critical coupling, $1.540(5)$ times the
mean-field value, corresponds---after rescaling the $|\bm{S}_i|^2 = 2$
normalization to unit spins---to $T_c = 1.299(4)$.
This precise reference value provides quantitative grounding for
dual theories of quantum spin liquids and multipolar ordered materials
where the isotropic diamond-lattice $XY$ model serves as the effective
description.\cite{Watanabe2026}
\begin{acknowledgments}
The work of S.W. is supported by JST SPRING, Grant No.~JPMJSP2108.
The work of Y.M. is supported by JSPS KAKENHI Grant No.~JP25H01247.
The work of H.W. is supported by JSPS KAKENHI Grant No.~JP24K00541.
\end{acknowledgments}

\end{document}